\documentclass[onecollarge,natbib]{svjour2}
\bibpunct{[}{]}{;}{n}{}{,} % to get "[numbered]" references from natbib
\smartqed  % flush right qed marks, e.g. at end of proof
\usepackage{graphicx}
\journalname{Few Body Systems}
\usepackage{pzccal}

\newcommand{\lsim}{\mathrel{\rlap{\lower4pt\hbox{\hskip0pt$\sim$}}
\raise1pt\hbox{$<$}}}           %less than or approx. symbol
\newcommand{\gsim}{\mathrel{\rlap{\lower4pt\hbox{\hskip0pt$\sim$}}
\raise1pt\hbox{$>$}}}           %greater than or approx. symbol

\usepackage{color}
\definecolor{purple}{rgb}{0.5,0,0.5}
\definecolor{blue}{rgb}{0.0,0,0.9}

\begin{document}

\title{Looking into the matter of light-quark hadrons\thanks{Presented by C.\,D.~Roberts at LIGHTCONE 2011, 23 - 27 May, 2011, Dallas.}
}

\authorrunning{Craig D.~Roberts}
\titlerunning{Looking into the matter of light-quark hadrons}        % if too long for running head
\sloppy

\author{Craig D.~Roberts
}


\institute{Craig D.~Roberts  (\email{cdroberts@anl.gov})
\at
Physics Division, Argonne National Laboratory, Argonne, Illinois 60439, USA;
\at
Department of Physics, Center for High Energy Physics and State Key Laboratory of Nuclear Physics and Technology, Peking University, Beijing 100871, China;
\at
Department of Physics, Illinois Institute of Technology, Chicago, Illinois 60616-3793, USA.
}

\date{Version: 28 September 2011}
%\date{Version: 22 September 2011}
%\date{Version: 16 September 2011}

\maketitle

\begin{abstract}
In tackling QCD, a constructive feedback between theory and extant and forthcoming experiments is necessary in order to place constraints on the infrared behaviour of QCD's $\beta$-function, a key nonperturbative quantity in hadron physics.  The Dyson-Schwinger equations provide a tool with which to work toward this goal.  They connect confinement with dynamical chiral symmetry breaking, both with the observable properties of hadrons, and hence provide a means of elucidating the material content of real-world QCD.  This contribution illustrates these points via comments on: in-hadron condensates; dressed-quark anomalous chromo- and electro-magnetic moments; the spectra of mesons and baryons, and the critical role played by hadron-hadron interactions in producing these spectra.
\end{abstract}

%\tableofcontents

\section{Introduction}
\label{intro}
Solving QCD presents a fundamental problem that is unique in the history of science.  We have never before been confronted by a theory whose elementary excitations are not those degrees-of-freedom readily accessible via experiment; i.e., whose elementary excitations are \emph{confined}.  Moreover, QCD generates forces which are so strong that less-than 2\% of a nucleon's mass can be attributed to the so-called current-quark masses that appear in QCD's Lagrangian; viz., forces capable of generating mass \emph{from nothing}, a phenomenon known as dynamical chiral symmetry breaking (DCSB).

Neither confinement nor DCSB is apparent in QCD's Lagrangian and yet they play the dominant role in determining the observable characteristics of real-world QCD.  The physics of hadronic matter is ruled by \emph{emergent phenomena}, such as these, which can only be elucidated and understood through the use of nonperturbative methods in quantum field theory.  This is both the greatest novelty and the greatest challenge within the Standard Model.  We must find essentially new ways and means to explain precisely via mathematics the observable content of QCD.  Building a bridge between QCD and the observed properties of hadrons is one of the key problems for modern science.

The complex of Dyson-Schwinger equations (DSEs) is a puissant tool, which has been employed with success to study confinement and DCSB, and their impact on hadron observables.  In this contribution, I will briefly describe selected recent highlights, some of which are covered more fully in recent reviews \cite{Roberts:2007jh,Roberts:2007ji,Holt:2010vj,Chang:2011vu}.

\section{Confinement}
It is worth stating plainly that the potential between infinitely-heavy quarks measured in simulations of quenched lattice-QCD -- the so-called static potential -- is \emph{irrelevant} to the question of confinement in the real world, in which light quarks are ubiquitous.  In fact, it is a basic feature of QCD that light-particle creation and annihilation effects are essentially nonperturbative.  It is therefore impossible in principle to compute a potential between two light quarks \cite{Bali:2005fu,Chang:2009ae}.  Hence, in discussing this physics, linearly rising potentials, flux-tube models, etc., have no connection with nor justification via QCD.

\begin{figure}[t]
\vspace*{7ex}

\hspace*{-1ex}\begin{minipage}[t]{1.0\textwidth}
\begin{minipage}[t]{1.0\textwidth}
\leftline{\includegraphics[clip,width=0.53\textwidth]{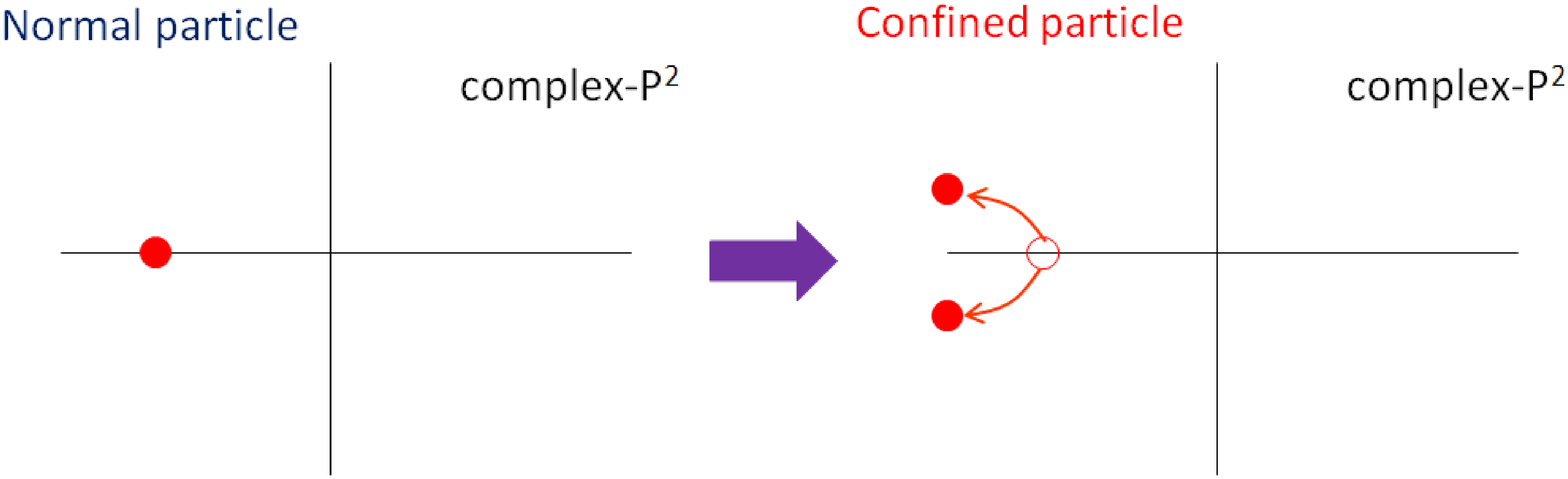}}
\end{minipage}\vspace*{-26ex}

\begin{minipage}[t]{1.0\textwidth}
\rightline{\includegraphics[clip,width=0.45\textwidth]{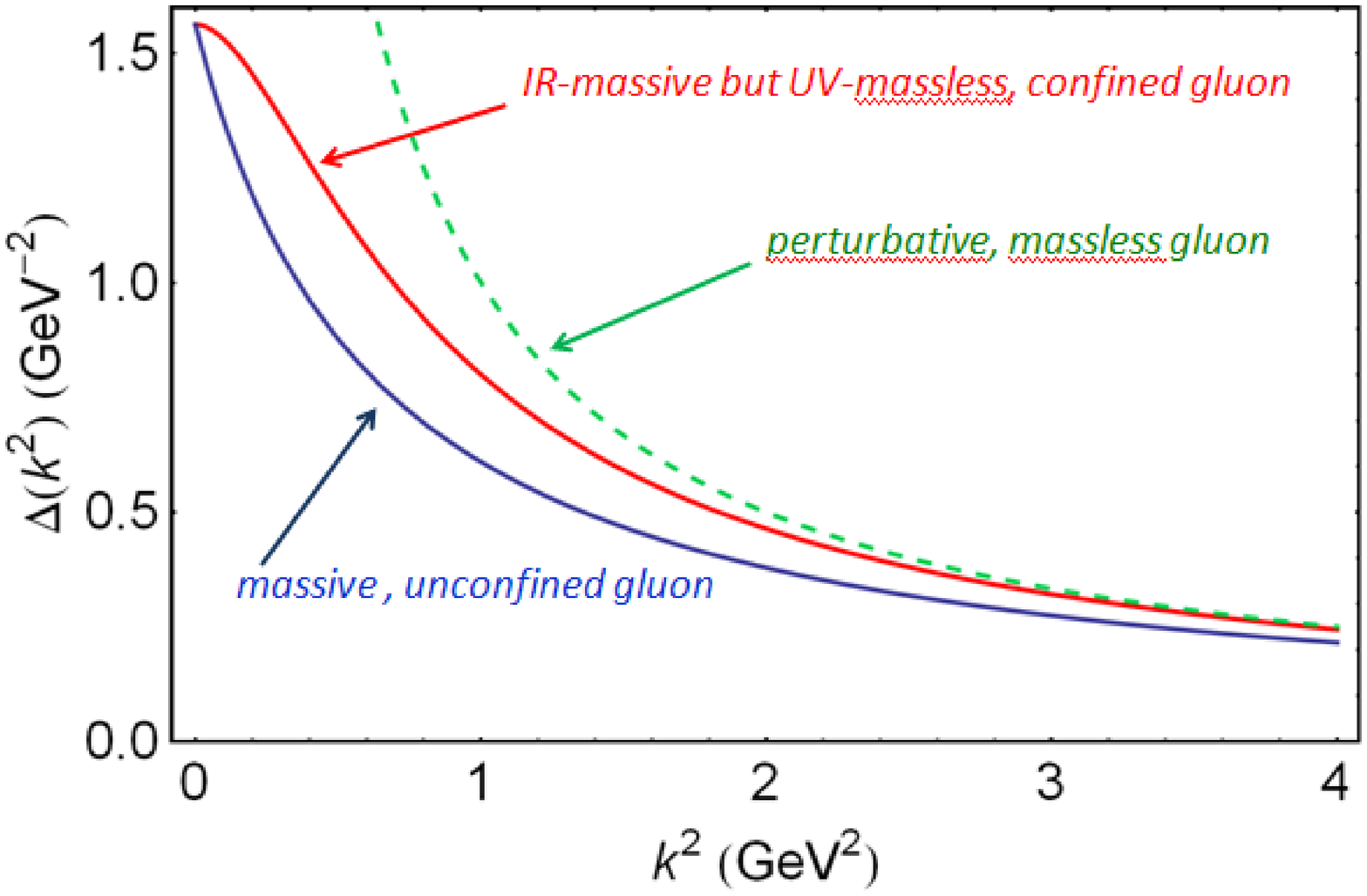}}
\end{minipage}\vspace*{3ex}
\end{minipage}
\caption{\label{fig:confined} \emph{Left panel} -- An observable particle is associated with a pole at timelike-$P^2$, which becomes a branch point if, e.g., the particle is dressed by photons.
\emph{Middle panel} -- When the dressing interaction is confining, the real-axis mass-pole splits, moving into pairs of complex conjugate poles or branch points.  No mass-shell can be associated with a particle whose propagator exhibits such singularity structure.
\emph{Right panel} -- $\Delta(k^2)$, the function that describes dressing of a Landau-gauge gluon propagator, plotted for three distinct cases.
A bare gluon is described by $\Delta(k^2) = 1/k^2$ (the dashed line), which is convex on $k^2\in (0,\infty)$.  Such a propagator has a representation in terms of a non-negative spectral density.
In some theories, interactions generate a mass in the transverse part of the gauge-boson propagator, so that $\Delta(k^2) = 1/(k^2+m_g^2)$, which can also be represented in terms of a non-negative spectral density.
In QCD, however, self-interactions generate a momentum-dependent mass for the gluon, which is large at infrared momenta but vanishes in the ultraviolet \protect\cite{Rodriguez-Quintero:2011}.  This is illustrated by the curve labelled ``IR-massive but UV-massless.''  With the generation of a mass-\emph{function}, $\Delta(k^2)$ exhibits an inflexion point and hence cannot be expressed in terms of a non-negative spectral density.
}
\end{figure}

\begin{figure}[t]

\leftline{%\hspace*{2em}%
\includegraphics[clip,width=0.50\textwidth]{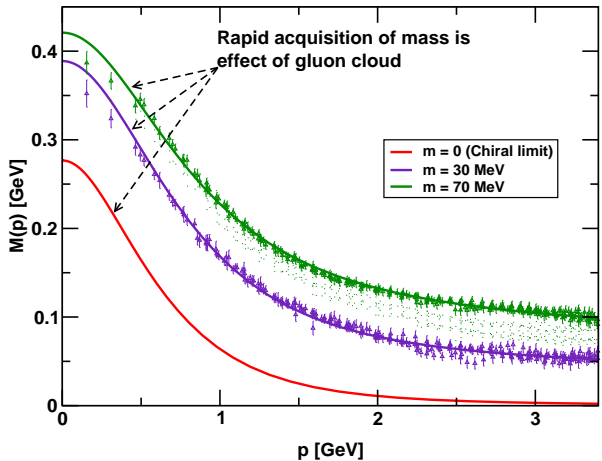}}\vspace*{-41ex}

\rightline{\parbox{0.42\textwidth}{
\caption{\label{fig:Mp} Dressed-quark mass function, $M(p)$: \emph{solid curves} -- DSE results, \protect\cite{Bhagwat:2006tu}, ``data'' -- lattice-QCD simulations \protect\cite{Bowman:2005vx}.  (NB.\ $m=70\,$MeV is the uppermost curve.  Current-quark mass decreases from top to bottom.)  %In QCD, the vast bulk of the light-quark constituent mass arises from gluons through DCSB.
The constituent mass arises from a cloud of low-momentum gluons attaching themselves to the current-quark: DCSB is a truly nonperturbative effect that generates a quark mass \emph{from nothing}; namely, it occurs even in the chiral limit, as evidenced by the $m=0$ curve}}}
\vspace*{11ex}

\end{figure}

On the other hand, confinement can be related to the analytic properties of QCD's Schwinger functions; i.e., the dressed-propagators and -vertices.  This perspective was laid out in Ref.\,\cite{Krein:1990sf}.  Whilst there is a great deal of mathematical background to this observation, it is readily illustrated, Fig.\,\ref{fig:confined}.  The simple pole of an observable particle produces a propagator that is a monotonically-decreasing convex function, whereas the evolution depicted in the middle-panel of Fig.\,\ref{fig:confined} is manifest in the propagator as the appearance of an inflexion point at $P^2 > 0$.  To complete the illustration, consider $\Delta(k^2)$, which is the single scalar function that describes the dressing of a Landau-gauge gluon propagator.  Three possibilities are exposed in the right-panel of Fig.\,\ref{fig:confined}.  The inflexion point possessed by $M(p^2)$, visible in Fig.\,\ref{fig:Mp}, entails, too, that the dressed-quark is confined.

From the perspective that confinement is related to the analytic properties of QCD's Schwinger functions, the question of light-quark confinement may be translated into the challenge of charting the infrared behavior of QCD's $\beta$-function. (The behavior of the $\beta$-function on the perturbative domain is well known.)  This is a well-posed problem whose solution is a primary goal of hadron physics; e.g., Refs.\,\cite{Qin:2011dd,Brodsky:2010ur,Aguilar:2010gm}.  It is the $\beta$-function that is responsible for the behavior evident in Figs.\,\ref{fig:confined} and \ref{fig:Mp}, and thereby the scale-dependence of the structure and interactions of dressed-gluons and -quarks.  One of the more interesting of contemporary questions is whether it is possible to reconstruct the $\beta$-function, or at least constrain it tightly, given empirical information on the gluon and quark mass functions.

\begin{figure}[t]
\leftline{\includegraphics[clip,width=0.48\textwidth]{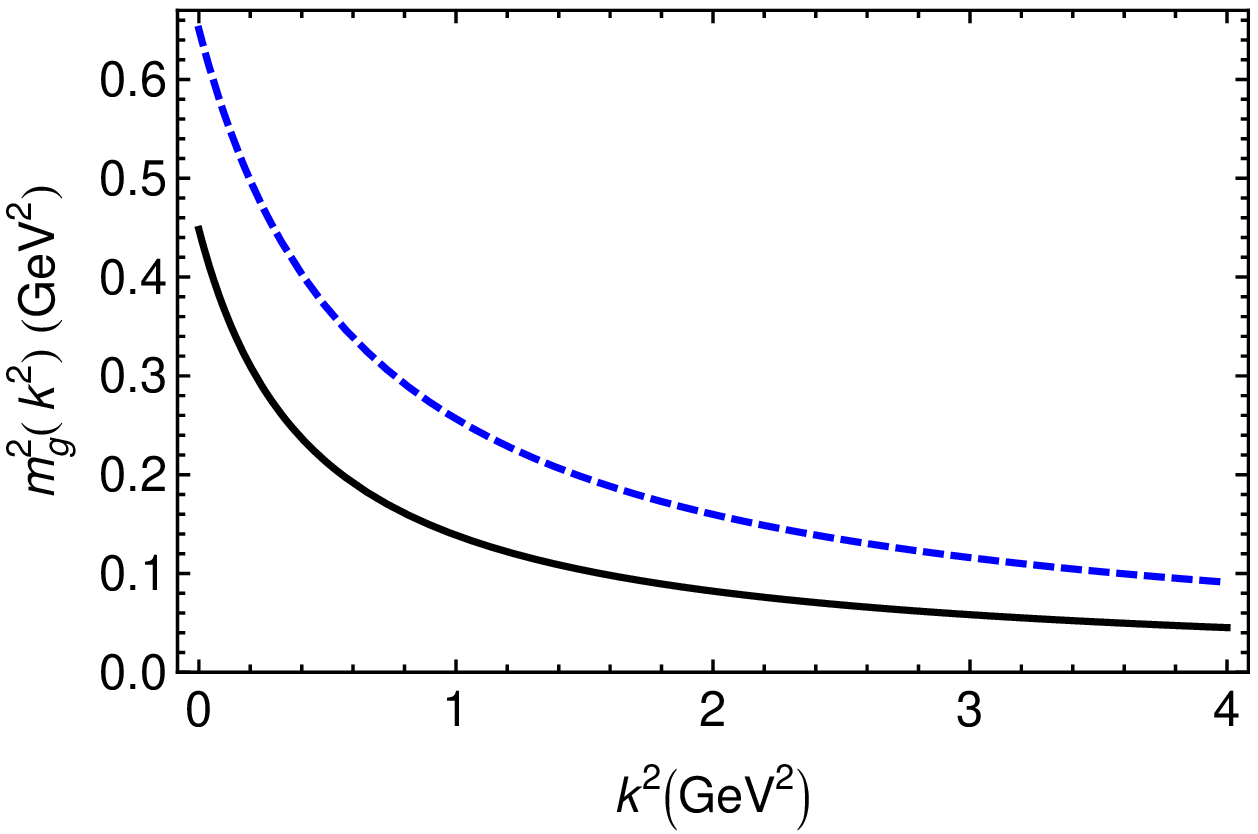}}
\vspace*{-36.2ex}

\rightline{\includegraphics[clip,width=0.49\textwidth]{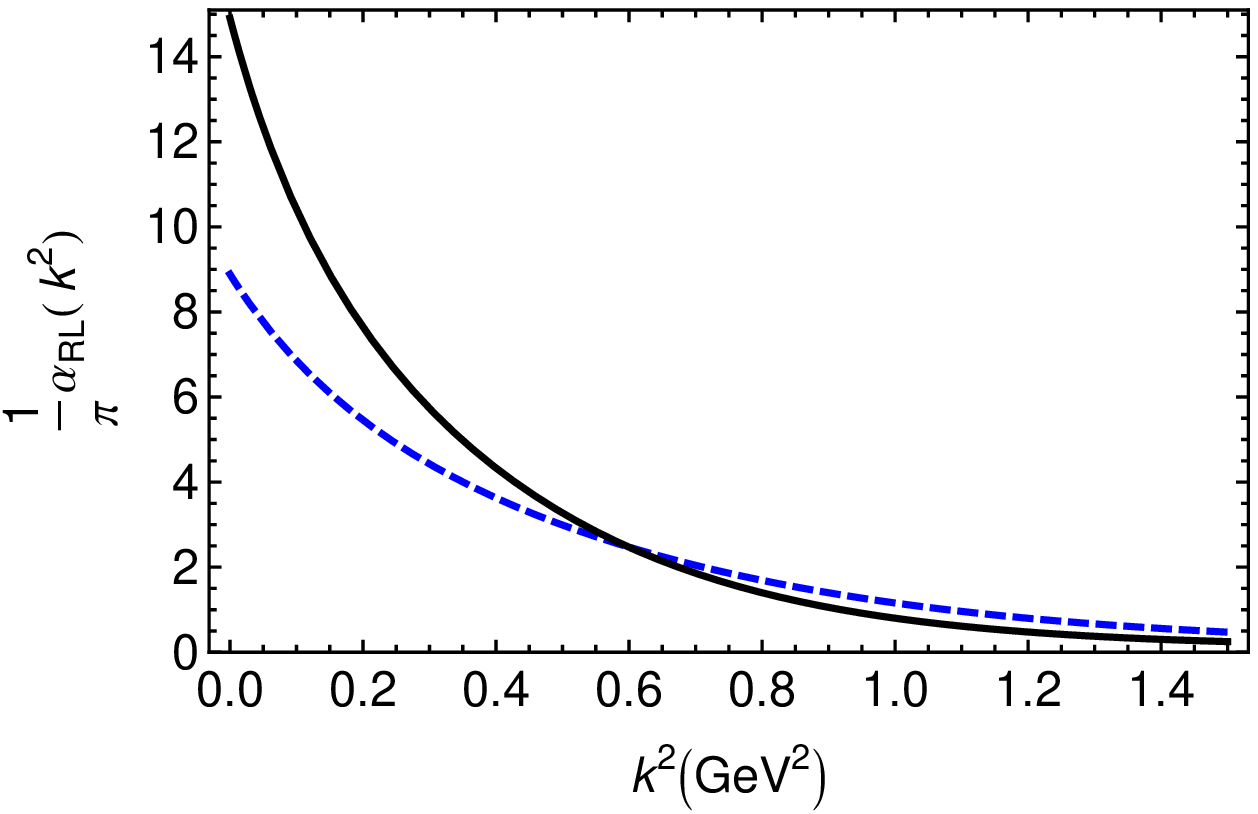}}

\caption{\label{alphaeff}
\emph{Left panel} -- Rainbow-ladder gluon running-mass; and
\emph{right panel} -- rainbow-ladder effective running-coupling, both determined in a DSE analysis of properties of light-quark mesons.  The dashed curves illustrate forms for these quantities that provide the more realistic picture \protect\cite{Qin:2011dd,Qin:2011xq}.  (Figures drawn from Ref.\,\protect\cite{Qin:2011dd}.)
}
\end{figure}

Experiment-theory feedback shows promise for providing the latter \cite{Roberts:2011rr,Gothe:2011up,MokeevI}.  This is illustrated through Fig.\,\ref{alphaeff}, which depicts the running-gluon-mass, analogous to $M(p)$ in Fig.\,\ref{fig:Mp}, and the running-coupling determined by analysing a range of properties of light-quark ground-state, radially-excited and exotic scalar-, vector- and flavoured-pseudoscalar-mesons in the rainbow-ladder truncation, which is leading order in a symmetry-preserving DSE truncation scheme \cite{Bender:1996bb}.  Consonant with modern DSE- and lattice-QCD results \cite{Rodriguez-Quintero:2011}, these functions derive from a gluon propagator that is a bounded, regular function of spacelike momenta, which achieves its maximum value on this domain at $k^2=0$ \cite{Aguilar:2010gm,Bowman:2004jm,Aguilar:2009nf}, and a dressed-quark-gluon vertex that does not possess any structure which can qualitatively alter this behaviour \cite{Skullerud:2003qu,Bhagwat:2004kj}.  In fact, the dressed-gluon mass drawn here produces a gluon propagator much like the curve labelled ``IR-massive but UV-massless'' in the right-panel of Fig.\,\ref{fig:confined}.

Notably, the value of $M_g=m_g(0)\sim 0.7\,$GeV is typical \cite{Bowman:2004jm,Aguilar:2009nf}; and the infrared value of the coupling, $\alpha_{RL}(M_g^2)/\pi = 2.2$, is interesting because a context is readily provided.
With nonperturbatively-massless gauge bosons, the coupling below which DCSB breaking is impossible via the gap equations in QED and QCD is $\alpha_c/\pi \approx 1/3$ \cite{Roberts:1989mj,Bloch:2002eq,Bashir:1994az}.
In a symmetry-preserving regularisation of a vector$\,\times\,$vector contact-interaction used in rainbow-ladder truncation, $\alpha_c/\pi \approx 0.4$; and a description of hadron phenomena requires $\alpha/\pi \approx 1$ \cite{Roberts:2011wy}.
With nonperturbatively massive gluons and quarks, whose masses and couplings run, the infrared strength required to describe hadron phenomena in rainbow-ladder truncation is unsurprisingly a little larger.
%--lattice inferred m=0.6 alpha_Q-latt/Pi=2
Moreover, whilst a direct comparison between $\alpha_{RL}$ and a coupling, $\alpha_{QLat}$, inferred from quenched-lattice results is not sensible, it is nonetheless curious that $\alpha_{QLat}(0)\lsim\alpha_{RL}(0)$ \cite{Aguilar:2010gm}.
It is thus noteworthy that with a more sophisticated, nonperturbative DSE truncation \cite{Chang:2009zb,Chang:2011ei}, some of the infrared strength in the gap equation's kernel is shifted from the gluon propagator into the dressed-quark-gluon vertex.  This cannot materially affect the net infrared strength required to explain observables but does reduce the amount attributed to the effective coupling.

\section{Dynamical chiral symmetry breaking}
Whilst the nature of confinement is still debated, Fig.\,\ref{fig:Mp} shows that DCSB is a fact.  It is the most important mass generating mechanism for visible matter in the Universe, being responsible for roughly 98\% of the proton's mass \cite{Flambaum:2005kc}.  Indeed, the Higgs mechanism is (almost) irrelevant to light-quarks.  In Fig.\,\ref{fig:Mp} one observes the current-quark of perturbative QCD evolving into a constituent-quark as its momentum becomes smaller.  This behavior, and that illustrated in Figs.\,\ref{fig:confined}, has a marked influence, e.g., on the $Q^2$-dependence of hadron elastic and transition form factors, whose measurement is a large part of the programme planned at the upgraded JLab facility.  Therefore, in combination with well-constrained QCD-based theory, such data can potentially be used to chart the evolution of the mass function on $0.3 \lsim p/{\rm GeV} \lsim 1.2$, which is a domain that bridges the gap between nonperturbative and perturbative QCD.  This can assist in unfolding the relationship between confinement and DCSB.

The appearance of running masses for gluons and quarks is a quantum field theoretic effect, unrealisable in quantum mechanics.  It entails, moreover, that: quarks are not Dirac particles; and the coupling between quarks and gluons involves structures that cannot be computed in perturbation theory.  Recent progress with the two-body problem in quantum field theory \cite{Chang:2009zb} has enabled these facts to be established \cite{Chang:2010hb}.  One may now plausibly argue that theory is in a  position to produce the first reliable symmetry-preserving, Poincar\'e-invariant prediction of the light-quark hadron spectrum \cite{Chang:2011ei}.

\section{Condensates are confined within hadrons}
The connection between DCSB and the generation of hadron masses was first described in Ref.\,\cite{Nambu:1961tp}, wherein it was represented as a vacuum phenomenon.  Two inequivalent classes of ground-state were identified in the mean-field treatment of a meson-nucleon field theory: symmetry preserving (Wigner phase); and symmetry breaking (Nambu phase).  In the symmetry breaking class, each of an uncountable infinity of distinct configurations is related to every other by a chiral rotation.  This is arguably the origin of the concept that strongly-interacting quantum field theories possess a nontrivial vacuum.

With the introduction of the parton model for the description of deep inelastic scattering (DIS), this notion was challenged via an argument \cite{Casher:1974xd} that DCSB can be realised as an intrinsic property of hadrons, instead of through a nontrivial vacuum exterior to the observable degrees of freedom.  This perspective is tenable because the essential ingredient required for dynamical symmetry breaking in a composite system is the existence of a divergent number of constituents and DIS provided evidence for the existence within every hadron of a sea of low-momentum partons.  This view has, however, received scant attention.  On the contrary, the introduction of QCD sum rules as a method to estimate nonperturbative strong-interaction matrix elements entrenched the belief that the QCD vacuum is characterised by numerous distinct, spacetime-independent condensates.

Notwithstanding the prevalence of this belief, it does lead to problems; e.g., entailing a cosmological constant that is $10^{46}$-times greater than that which is observed \cite{Turner:2001yu,Brodsky:2009zd}.  This unwelcome consequence is partly responsible for reconsideration of the possibility that the so-called vacuum condensates are in fact an intrinsic property of hadrons.  Namely, in a confining theory, condensates are not constant, physical mass-scales that fill all spacetime; instead, they are merely mass-dimensioned parameters that serve a practical purpose in some theoretical truncation schemes but otherwise do not have an existence independent of hadrons \cite{Brodsky:2009zd,Burkardt:1998dd,Brodsky:2008be,Brodsky:2010xf,Glazek:2011vg}.

Regarding the quark condensate, this perspective was recently elucidated for light pseudoscalar mesons \cite{Brodsky:2010xf}; and subsequently extended to all hadrons \cite{Chang:2011mu} through analysis enabled by two key realisations.  The first is connected with the in-pseudoscalar-meson quark condensate, which is a quantity with an exact expression in QCD; viz. \cite{Maris:1997hd,Maris:1997tm}, $\kappa^\zeta_{P_{f_1 f_2}} = \rho_{P_{f_1 f_2}}^\zeta f_{P_{f_1 f_2}}$, where ($k_\pm = k\pm P/2$)
\begin{eqnarray}
\label{fpigen}
i f_{P_{f_1 f_2}} K_\mu = \langle 0 | \bar q_{f_2} \gamma_5 \gamma_\mu q_{f_1} |P \rangle &=&  Z_2(\zeta,\Lambda)\; {\rm tr}_{\rm CD}
\int_k^\Lambda i\gamma_5\gamma_\mu S_{f_1}(k_+) \Gamma_{P_{f_1 f_2}}(k;K) S_{f_2}(k_-)\,, \\
i\rho_{P_{f_1 f_2}}^\zeta = -\langle 0 | \bar f_2 i\gamma_5 f_1 |P \rangle
&=& Z_4(\zeta,\Lambda)\; {\rm tr}_{\rm CD}
\int_k^\Lambda \gamma_5 S_{f_1}(k_+) \Gamma_{P_{f_1 f_2}}(k;K) S_{f_2}(k_-) \,,\label{rhogen}\\
\label{GMORP}
f_{P_{f_1 f_2}}^2 m_{P_{f_1 f_2}}^2 &=& [m_{f_1}^\zeta +m_{f_2}^\zeta]\, \kappa^\zeta_{P_{f_1 f_2}}.
%
%\label{fpigen}
%
\end{eqnarray}
Here $\int_k^\Lambda:=\int^\Lambda \!\! \mbox{\footnotesize $\frac{d^4 \ell}{(2\pi)^4}$}$ is a Poincar\'e-invariant regularization of the integral, with $\Lambda$ the ultraviolet regularization mass-scale, $Z_{2,4}$ are renormalisation constants, with $\zeta$ the renormalisation point, $\Gamma_{P_{f_1 f_2}}$ is the meson's Bethe-Salpeter amplitude, and $S_{f_1,f_2}$ are the component dressed-quark propagators, with $m_{f_1,f_2}$ the associated current-quark masses.  The quantity in Eq.\,(\ref{fpigen}) is the pseudoscalar meson's leptonic decay constant; i.e., the pseudovector projection of the meson's Bethe-Salpeter wave-function onto the origin in configuration space; that in Eq.\,(\ref{rhogen}) is its pseudsocalar analogue; and $m_{P_{f_1 f_2}}$ is the meson's mass.  The initial step in extending the concept was a proof that the in-pseudoscalar-meson quark condensate, $\kappa^\zeta_{P_{f_1 f_2}}$, can rigorously be represented through the pseudoscalar-meson's scalar form factor at zero momentum transfer, $Q^2=0$; viz., ($m^\zeta_{f_1 f_2}=m^\zeta_{f_1}+ m^\zeta_{f_2}$)
\begin{equation}
{\cal S}_{P_{f_1 f_2}} := -\langle P_{f_1 f_2}|\mbox{\small $\frac{1}{2}$} (\bar q_{f_1} q_{f_2}+\bar q_{f_1} q_{f_2}) | P_{f_1 f_2} \rangle
=\frac{\partial}{\partial m_{f_1 f_2}^\zeta}m^2_{P_{f_1 f_2}}
=
\frac{ \kappa^\zeta_{P_{f_1 f_2}} }{f^2_{P_{f_1 f_2}}} + m_{f_1 f_2}^\zeta
\frac{\partial}{\partial m_{f_1 f_2}^\zeta}  \left[ \frac{ \kappa^\zeta_{P_{f_1 f_2}} }{f^2_{P_{f_1 f_2}}}\right].
\end{equation}

\begin{figure}[t]
%\vspace*{-2ex}

\leftline{\includegraphics[clip,width=0.25\textwidth]{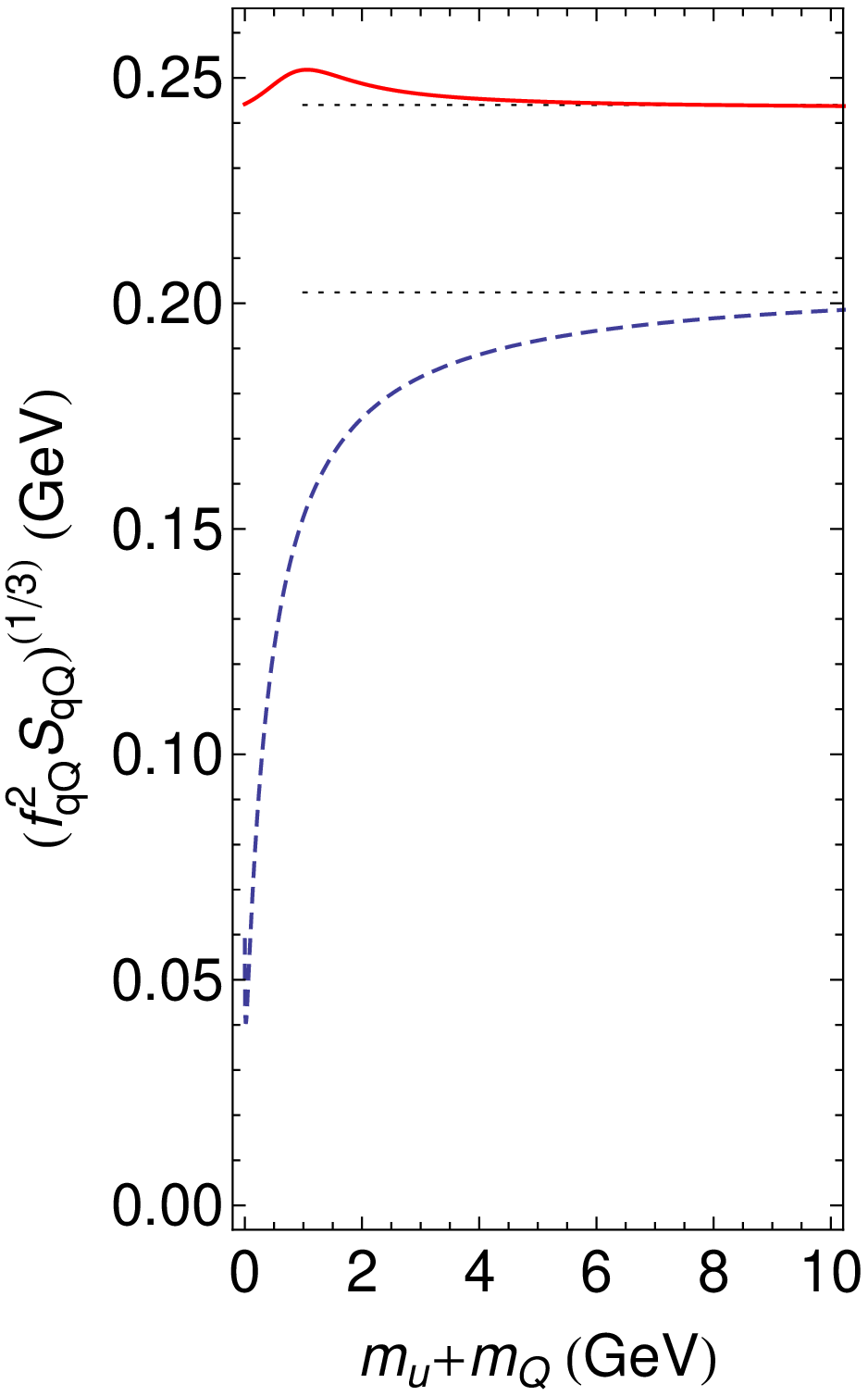}}
\vspace*{-46.3ex}

\hspace*{12.0em}{\includegraphics[clip,width=0.255\textwidth]{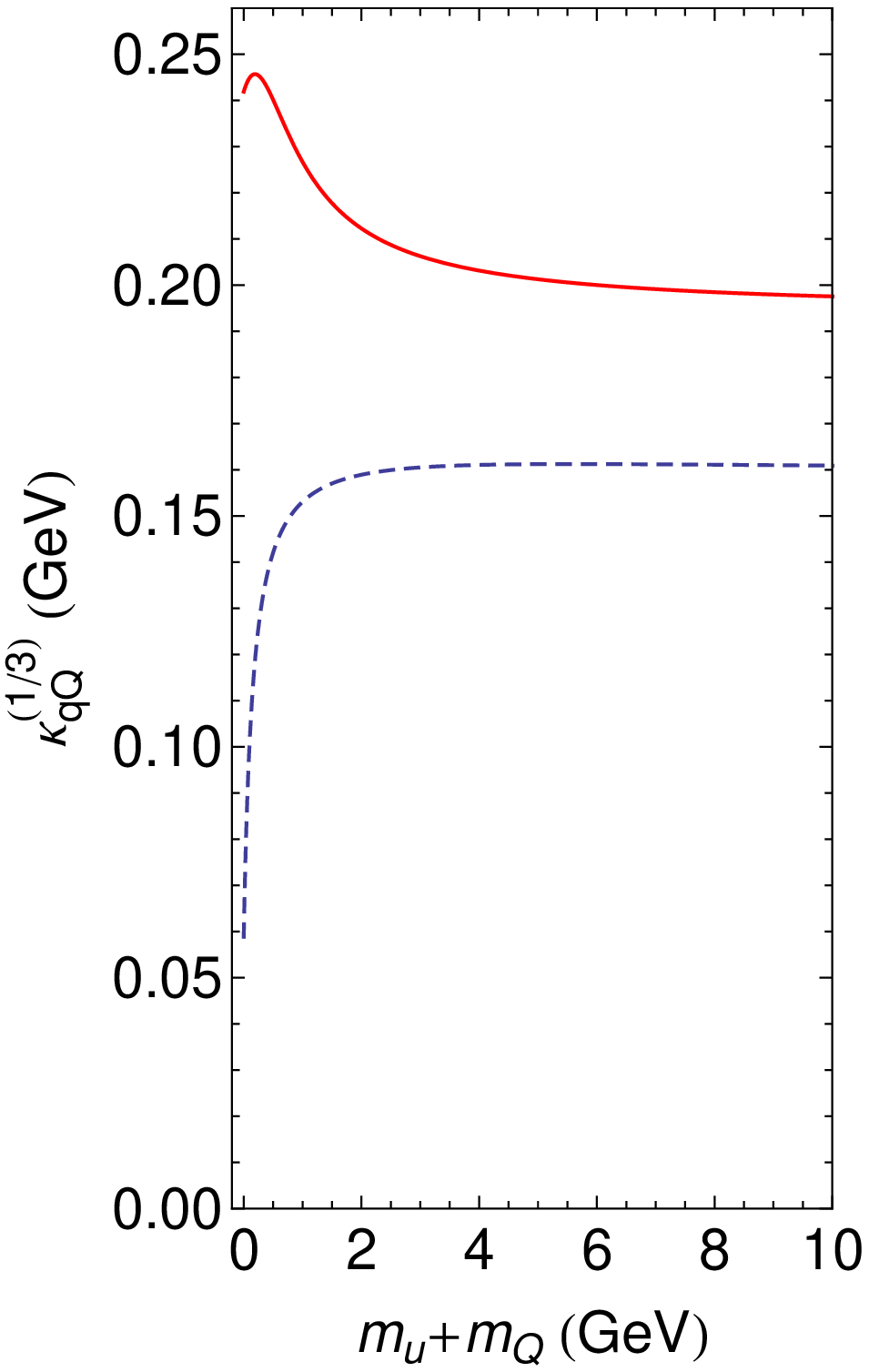}}
\vspace*{-46.3ex}

\hspace*{24.1em}{\includegraphics[clip,width=0.25\textwidth]{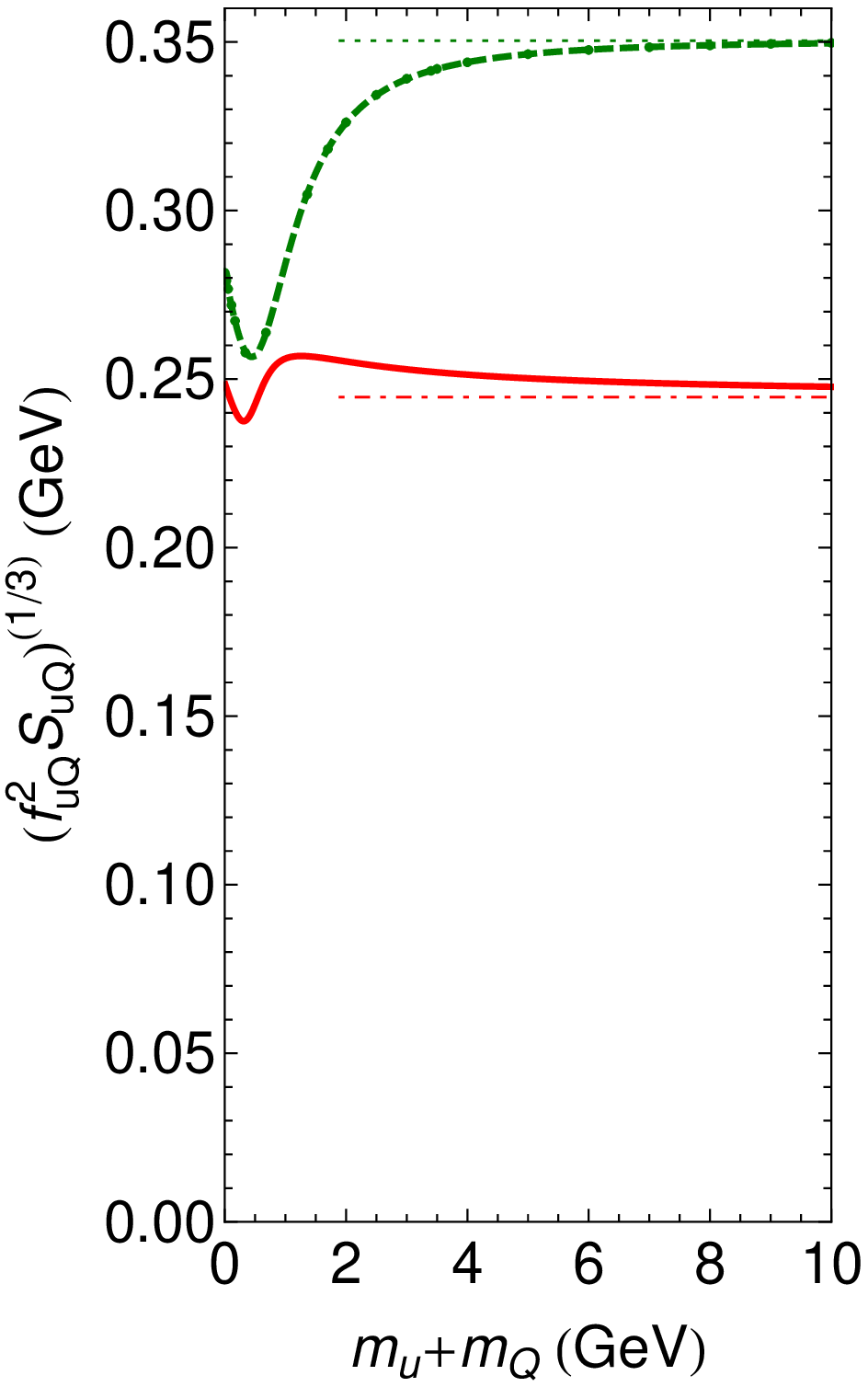}}
\vspace*{-46.3ex}

\rightline{\includegraphics[clip,width=0.25\textwidth]{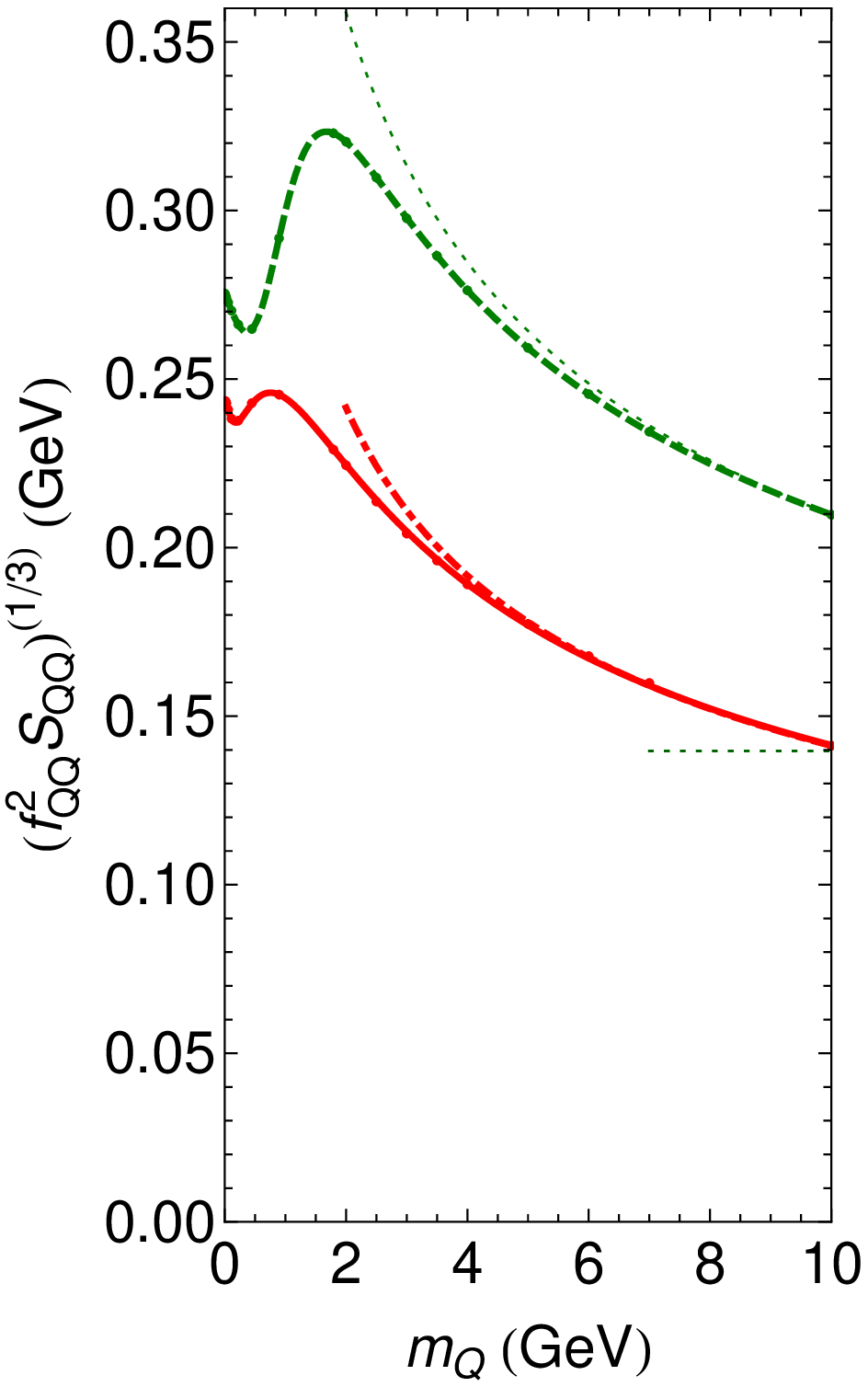}}

\caption{\label{Fig2H}
\underline{Heavy-light systems}.
\emph{Panel One} -- solid curve, $[f^2_{P_{qQ}} {\cal S}_{P_{qQ}}]^{1/3}$; dashed curve, $[f^2_{S_{qQ}} {\cal S}_{S_{qQ}}]^{1/3}$; and dotted lines, heavy-quark-limit values of $[2 \kappa_{qQ}]^{1/3}$ computed directly from Eqs.\,(\ref{fpigen}), (\ref{rhogen}) and Eqs.\,(\ref{fsigmagen}), (\ref{rhosigmagen}), respectively.
\emph{Panel Two} -- $[\kappa_{qQ}]^{1/3}$ computed directly from Eqs.\,(\ref{fpigen}),(\ref{rhogen}) (pseudoscalar, solid curve) and Eqs.\,(\ref{fsigmagen}), (\ref{rhosigmagen}) (scalar, dashed curve).
\emph{Panel Three} -- solid curve, $[f^2_{P_{qQ}} {\cal S}_{P_{qQ}}]^{1/3}$; dashed curve, $[f^2_{V_{qQ}} {\cal S}_{V_{qQ}}]^{1/3}$; and dash-dotted and dotted lines, respectively, the $m_Q\to \infty$ values.
\underline{Heavy-heavy systems}.
\emph{Panel Four} -- solid curve, $[f^2_{P_{\bar QQ}} {\cal S}_{P_{\bar QQ}}]^{1/3}$; and dashed curve, $[f^2_{V_{\bar QQ}} {\cal S}_{V_{\bar QQ}}]^{1/3}$.
The dot-dashed and dotted curves show the $m_{\bar Q} = m_Q \to \infty$ limiting behaviours: $f^2_{{\bar QQ}} {\cal S}_{{\bar QQ}}\propto 1/m_Q$; and the short horizontal dashed line indicates the value of $2 \kappa_{P_{\bar QQ}}$ at $m_{\bar Q}=m_Q=10\,$GeV.
These four panels: illustrate that $\chi:=f^2 {\cal S}$ is a smoothly varying measure of DCSB for pseudoscalar-, scalar- and vector-mesons, with values typical of those usually associated with this phenomenon; and confirm Eqs.\,(\protect\ref{HQlimits}).
(NB.\ All results computed using the symmetry-preserving regularisation of a vector$\,\times\,$vector contact-interaction described in Ref.\,\protect\cite{Roberts:2011wy,Roberts:2011cf}: $m_u$ is fixed at 7\,MeV; and the chiral-limit value of the in-pseudoscalar-meson condensate is $\kappa^0 = (0.24\,$GeV$)^3$.)
}
\end{figure}

The second step employs a mass formula for scalar mesons, exact in QCD; viz. \cite{Chang:2011mu},
\begin{eqnarray}
\label{gmorS}
f_{S_{f_1 f_2}} m_{S_{f_1 f_2}}^2 &=& -[m_{f_1}^\zeta - m_{f_2}^\zeta] \, \rho_{S_{12}}^\zeta,\\
f_{S_{f_1 f_2}} K_\mu
& = & Z_2\, {\rm tr}_{\rm CD}\!\!\!
\int_k^\Lambda i\gamma_\mu S_{f_1}(k_+) \Gamma_{S_{f_1 f_2}}(k;K) S_{f_2}(k_-)\,, \rule{2em}{0ex}
\label{fsigmagen} \\
\rho^\zeta_{S_{f_1 f_2}}
& = & - Z_4\, {\rm tr}_{\rm CD}\!\!\!
\int_k^\Lambda S_{f_1}(k_+) \Gamma_{S_{f_1 f_2}}(k;K) S_{f_2}(k_-) , \rule{2em}{0ex}\label{rhosigmagen}
\end{eqnarray}
which was used to prove that the in-scalar-meson quark condensate is, analogously and rigorously, connected with the scalar-meson's scalar form factor at zero momentum transfer.  Moreover the following limiting cases were also established:
\begin{equation}
\label{HQlimits}
f_{P,S}^2 \, {\cal S}^\zeta_{P,S} \stackrel{\rm chiral\,limit}{=} \kappa^0 = -\langle \bar q q \rangle^0
\; \mbox{and}\;
f_{P,S}^2 \, {\cal S}^\zeta_{P,S} \stackrel{\rm heavy\,quark(s)}{=} 2 \kappa^\zeta_{P,S},
\end{equation}
where $\langle \bar q q \rangle^0$ is precisely the quantity that is widely known as the \emph{vacuum quark condensate} through any of its definitions \cite{Langfeld:2003ye}.

With appeal then to demonstrable results of heavy-quark symmetry in QCD, it was argued that the $Q^2=0$ values of vector- and pseudovector-meson scalar form factors also determine the in-hadron condensates in these cases, and that this expression for the concept of in-hadron quark condensates is readily extended to the case of baryons.  Thus, through the $Q^2=0$ value of any hadron's scalar form factor, one can extract the magnitude of a quark condensate in that hadron which is a reasonable and realistic measure of dynamical chiral symmetry breaking.

These ideas are illustrated in Fig.\,\ref{Fig2H}, using the symmetry-preserving regularisation of a vector$\,\times\,$vector contact interaction described in Refs.\,\cite{Roberts:2011wy,Roberts:2011cf}:
the left two panels describe evolution from the chiral limit to heavy-light pseudoscalar and scalar systems;
the third panel describes the same for heavy-light vector mesons and compares this with the kindred pseudoscalar's behaviour;
and the rightmost panel portrays evolution to heavy-heavy pseudoscalar- and vector-mesons.  %
It is notable that the in-vector-meson condensate is $(1.13)^3$-times greater than the in-pseudoscalar meson condensate in the neighbourhood of the chiral limit.

In the $m_Q\to \infty$ limit, this ratio rises to $(1.43)^3$, an outcome which finds explanation in the results: $m_{V_{qQ}}\approx m_{P_{qQ}}$ but $f^2_{V_{qQ}} \approx \mathpzc{f}^2_V/m_Q$, $f^2_{P_{qQ}} \approx \mathpzc{f}^2_P/m_Q$, with $\mathpzc{f}_V \neq \mathpzc{f}_P$.  In QCD, on the contrary \cite{Neubert:1993mb}: $\mathpzc{f}_V = \mathpzc{f}_P$, and hence the solid and dashed curves in this panel would merge.  In fact, based on the information in Refs.\,\cite{ElBennich:2010ha,Nakamura:2010zzi} one estimates $\kappa_B^{\bar m_b} = (0.60\pm 0.10\,{\rm GeV})^3$ and $\kappa_{B^\ast}^{\bar m_b} = (0.59^{+0.10}_{-0.05}\,{\rm GeV})^3$, values which may be compared with $\kappa_{P}^{0 \bar m_b} = (0.27\,{\rm GeV})^3$.
In the case of heavy-heavy mesons, the vector$\,\times\,$vector contact-interaction produces
$m_{V_{\bar Q Q}}^2 f^2_{V_{\bar QQ}}= m_{\bar Q Q} \kappa_{V_{\bar QQ}}=\,$constant$_V$,
$m_{P_{\bar Q Q}}^2 f^2_{P_{\bar QQ}}= m_{\bar Q Q} \kappa_{P_{\bar QQ}}=\,$constant$_P$,
where ``constant$_{V,P}$'' are independent of $m_{\bar Q Q}$.  In QCD, on the contrary, the result is $m_{{V,P}_{\bar Q Q}}^2 f^2_{{V,P}_{\bar QQ}}= m_{\bar Q Q} \kappa_{\bar QQ}=\,{\rm constant} \times m_{\bar Q Q}^4$, where ``constant'' is independent of $m_{\bar Q Q}$, and hence, again, one would compute curves that merge.

\section{Dressed-quark anomalous chromo- and electro-magnetic moments}
The appearance and behaviour of $M(p)$ in Fig.\,\ref{fig:Mp} are essentially quantum field theoretic effects, unrealisable in quantum mechanics.  The running mass connects the infrared and ultraviolet regimes of the theory, and establishes that the constituent-quark and current-quark masses are simply two connected points on a single curve separated by a large momentum interval.  QCD's dressed-quark behaves as a constituent-quark, a current-quark, or something in between, depending on the momentum of the probe which explores the bound-state containing the dressed-quark.  These remarks should make clear that QCD's dressed-quarks are not simply Dirac particles.

This is emphasised by a consideration of their anomalous magnetic moments.  In QCD the analogue of Schwinger's one-loop calculation can be performed to find an anomalous \emph{chromo}-magnetic moment for the quark.  There are two diagrams in this case: one similar in form to that in QED; and another owing to the gluon self-interaction.  One reads from Ref.\,\cite{Davydychev:2000rt} that the perturbative result vanishes in the chiral limit.  This is consonant with helicity conservation in perturbative massless-QCD, which entails that the quark-gluon vertex cannot perturbatively have a term with such helicity-flipping characteristics.  However, Fig.\,\ref{fig:Mp} demonstrates that chiral symmetry is dynamically broken strongly in QCD and one must therefore ask whether this affects the chromomagnetic moment.

Of course, it does, and it is now known that the effect is modulated by the strong momentum dependence of the dressed-quark mass-function \cite{Chang:2010hb}.  In fact, dressed-quarks possess a large, dynamically-generated anomalous chromomagnetic moment, which produces an equally large anomalous electromagnetic moment that has a material impact on nucleon magnetic form factors \cite{Chang:2011tx} and also very likely on nucleon transition form factors.  Furthermore, given the magnitude of the muon ``$g_\mu-2$ anomaly'' and its assumed importance as an harbinger of physics beyond the Standard Model \cite{Jegerlehner:2009ry}, it might also be worthwhile to make a quantitative estimate of the contribution to $g_\mu-2$ from the quark's DCSB-induced anomalous moments following, e.g., the computational pattern indicated in Ref.\,\cite{Goecke:2010if}.

\section{\mbox{\boldmath $a_1$}-\mbox{\boldmath $\rho$} mass splitting}
Following Ref.\,\cite{Chang:2009zb} it is now possible to construct a symmetry-preserving kernel for the Bethe-Salpeter equation given any form for the dressed-quark-gluon vertex, $\Gamma_\mu$.  Owing to the importance of symmetries in forming the spectrum of a quantum field theory, this is a pivotal advance.  One may now use all available, reliable information to construct the best possible vertex \emph{Ansatz}.  The preceding section illustrated that this enables one to incorporate crucial nonperturbative effects, which any finite sum of contributions is incapable of capturing, and thereby prove that DCSB generates material, momentum-dependent anomalous chromo- and electro-magnetic moments for dressed light-quarks.

\begin{figure}[t]
%\vspace*{-2ex}

\leftline{\includegraphics[clip,width=0.48\textwidth]{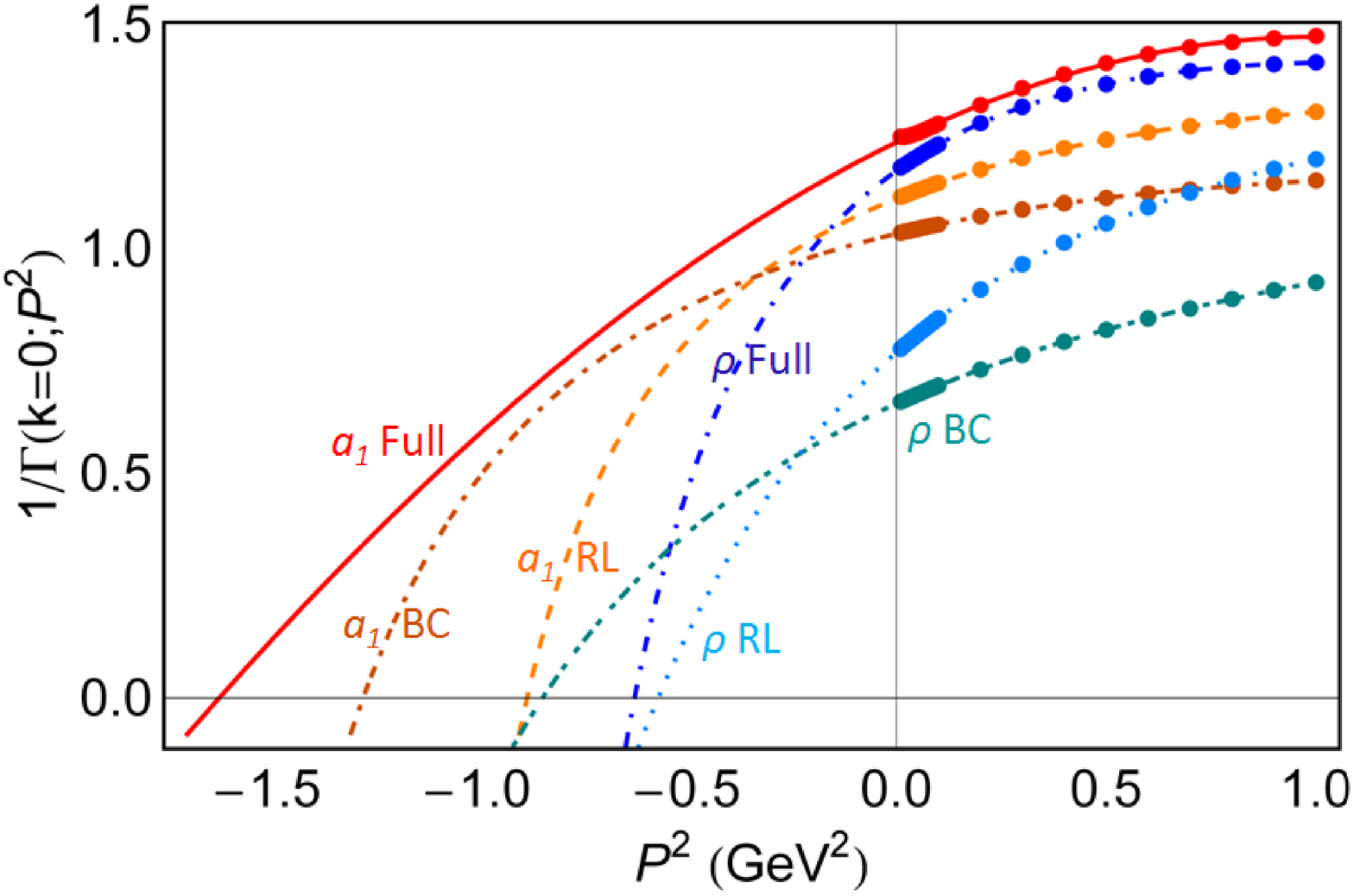}}
\vspace*{-38ex}

\rightline{\parbox{0.48\textwidth}{
\caption{\label{F1}
Illustration of a practical procedure for computing meson masses from information at spacelike momenta \protect\cite{Bhagwat:2007rj}.
%, which is analogous to that used in lattice-QCD.
%, which is fully described in Ref.\,\cite{Bhagwat:2007rj} and analogous to the method used in lattice-QCD.
\emph{Solid curve} -- $a_1$-meson, complete nonperturbative kernel; \emph{dot-dash-dash} -- $a_1$, kernel derived from the Ball-Chiu vertex (BC) \protect\cite{Ball:1980ay}; and \emph{dash} -- $a_1$, leading-order kernel (rainbow-ladder, RL).
\emph{Dot-dash curve} -- $\rho$-meson, complete nonperturbative kernel; \emph{Dot-dash-dot} -- $\rho$, BC-kernel; and \emph{dotted} -- $\rho$, RL-kernel.
\emph{Points} -- values of $1/\Gamma(k=0;P^2)$, where $\Gamma$ is the determined from the inhomogeneous vertex in the given channel computed with the kernel described.  Pad\'e approximants are constructed in each case; and the location of the zero is identified with $(-m_{\rm meson}^2)$.
(See Ref.\,\protect\cite{Chang:2011ei} for more details.)}}}
\medskip
\end{figure}

Reference~\protect\cite{Chang:2011ei} computes ground-state masses using the method detailed in Ref.\,\cite{Bhagwat:2007rj}, which ensures one need only solve the gap- and Bethe-Salpeter-equations at spacelike momenta.  This simplifies the numerical problem and is illustrated for the $\rho$- and $a_1$-channels in Fig.\,\ref{F1}.  More important in this figure, however, are the predicted masses for the mesons.  The $\rho$- and $a_1$-mesons have been known members of the spectrum for more than thirty years and are typically judged to be parity-partners; i.e., they would be degenerate if chiral symmetry were manifest in QCD.  Plainly, they are not, being split by more than $400\,$MeV (i.e., $> m_\rho/2$).  It is suspected that this large splitting owes to DCSB.  Hitherto, however, no symmetry-preserving treatment of bound-states could explain the splitting.  This is illustrated by the curves labelled ``RL,'' which show that whilst a good estimate of $m_\rho$ is readily obtained at leading-order in the systematic DSE truncation scheme of Ref.\,\cite{Bender:1996bb}, the axial-vector masses are much underestimated.  The defect persists at next-to-leading-order \cite{Watson:2004kd,Fischer:2009jm}.

The analysis in Ref.\,\cite{Chang:2011ei} points to a remedy for this longstanding failure.  Using the Poincar\'e-covariant, symmetry preserving formulation of the meson bound-state problem enabled by Ref.\,\cite{Chang:2009zb}, with nonperturbative kernels for the gap- and Bethe-Salpeter-equations, which incorporate effects of DCSB that are impossible to capture in any step-by-step procedure for improving upon the rainbow-ladder truncation, it provides realistic estimates of axial-vector meson masses.
In obtaining these results, Ref.\,\cite{Chang:2011ei} showed that the vertex \emph{Ansatz} used most widely in studies of DCSB, $\Gamma_\mu^{BC}$ \cite{Ball:1980ay}, is inadequate as a tool in hadron physics.  Used alone, it increases both $m_\rho$ and $m_{a_1}$ but yields $m_{a_1}-m_\rho=0.21\,$GeV, qualitatively unchanged from the rainbow-ladder result (see Fig.\,\ref{F1}).
A good description of axial-vector mesons is only achieved by including interactions derived from the dressed-quark anomalous chromomagnetic moment \cite{Chang:2010hb}.  This emphasises again that the dressed-quark-gluon and -quark-photon vertices involve structures that cannot be computed in perturbation theory.

\section{Unification of meson and baryon properties}
Owing to the importance of DCSB, full capitalisation on the results of forthcoming experimental programmes is only possible if the properties of meson and baryon ground- and excited-states can be correlated within a single, symmetry-preserving framework, where symmetry-preserving means that all relevant Ward-Takahashi identities are satisfied.  Constituent-quark-like models, which cannot incorporate the momentum-dependent dressed-quark mass-function, fail this test.  That is not to say, however, such models are worthless: they continue to be valuable because there is nothing that is yet providing a bigger picture.  Nevertheless, such models have no connection with quantum field theory and therefore not with QCD; and they are not ``symmetry-preserving'' and hence cannot veraciously connect meson and baryon properties.

An alternative is being pursued within quantum field theory via the Faddeev equation.  This analogue of the Bethe-Salpeter equation sums all possible interactions that can occur between three dressed-quarks.  A simplified equation \cite{Cahill:1988dx} is founded on the observation that an interaction which describes color-singlet mesons also generates nonpointlike quark-quark (diquark) correlations in the color-antitriplet channel \cite{Cahill:1987qr}.  The dominant correlations for ground state octet and decuplet baryons are scalar ($0^+$) and axial-vector ($1^+$) diquarks because, e.g., the associated mass-scales are smaller than the baryons' masses and their parity matches that of these baryons.  (Recent studies have confirmed the fidelity of the nonpointlike diquark approximation within the nucleon three-body problem; e.g. Ref.\,\protect\cite{Eichmann:2011vu}.) On the other hand, pseudoscalar ($0^-$) and vector ($1^-$) diquarks dominate in the parity-partners of those ground states \cite{Roberts:2011cf}.  This approach treats mesons and baryons on the same footing and, in particular, enables the impact of DCSB to be expressed in the prediction of baryon properties.

\begin{figure}[t]
\centerline{\includegraphics[clip,width=0.95\textwidth]{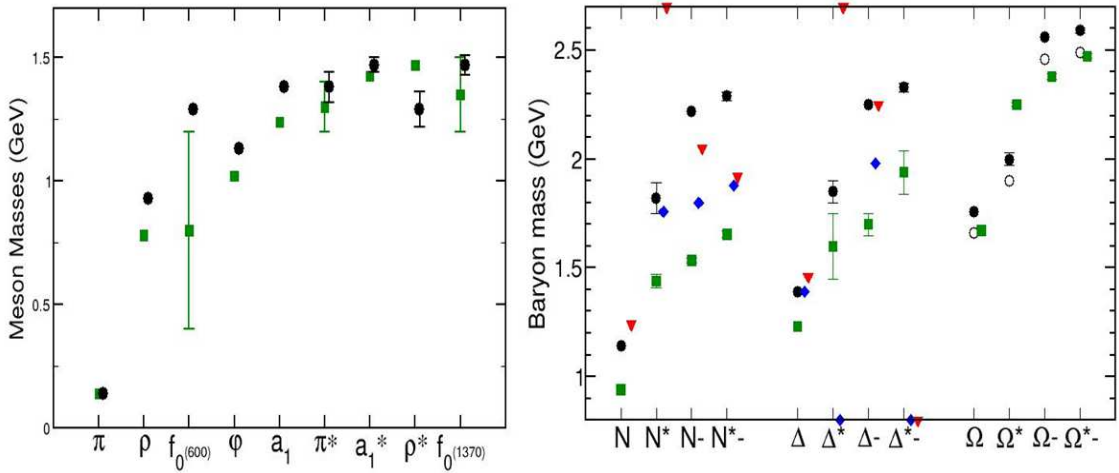}}
\caption{\label{Fig2}
Comparison between DSE-computed hadron masses (\emph{filled circles}) and: bare baryon masses from Ref.\,\protect\cite{Suzuki:2009nj} (\emph{filled diamonds}) and Ref.\,\protect\cite{Gasparyan:2003fp} (\emph{filled triangles}); and experiment \protect\cite{Nakamura:2010zzi}, \emph{filled-squares}.
For the coupled-channels models a symbol at the lower extremity indicates that no associated state is found in the analysis, whilst a symbol at the upper extremity indicates that the analysis reports a dynamically-generated resonance with no corresponding bare-baryon state.
In connection with $\Omega$-baryons the \emph{open-circles} represent a shift downward in the computed results by $100\,$MeV.  This is an estimate of the effect produced by pseudoscalar-meson loop corrections in $\Delta$-like systems at a $s$-quark current-mass.
}
\end{figure}

Building on lessons from meson studies \cite{Chang:2011vu}, a unified spectrum of $u,d$-quark hadrons was obtained using a symmetry-preserving regularization of a vector$\,\times\,$vector contact interaction \cite{Roberts:2011cf}.  That study simultaneously correlates the masses of meson and baryon ground- and excited-states within a single framework.  In comparison with relevant quantities, the computation produces $\overline{\mbox{rms}}$=13\%, where $\overline{\mbox{rms}}$ is the root-mean-square-relative-error$/$degree-of freedom.  As evident in Fig.\,\ref{Fig2}, the prediction uniformly overestimates the PDG values of meson and baryon masses \cite{Nakamura:2010zzi}.  Given that the employed truncation deliberately omitted meson-cloud effects in the Faddeev kernel, this is a good outcome, since inclusion of such contributions acts to reduce the computed masses.

\begin{figure}[t]
\leftline{\includegraphics[clip,width=0.52\textwidth]{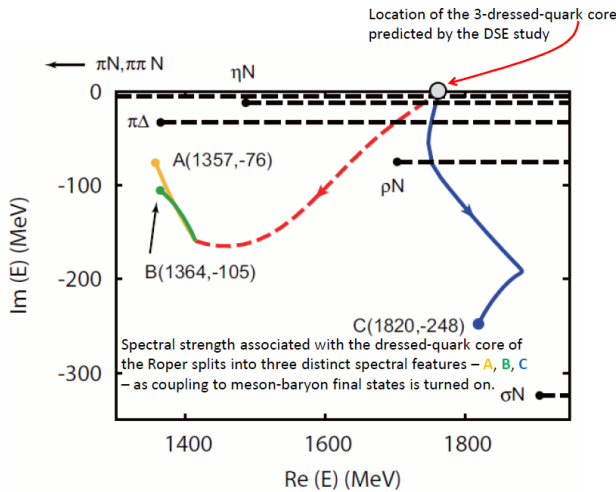}}
\vspace*{-37ex}

\rightline{\parbox{0.48\textwidth}{\caption{\label{ebac}
The Excited Baryon Analysis Center (EBAC) examined the $P_{11}$-channel and found that the two poles associated with the Roper resonance and the next higher resonance were all associated with the same seed dressed-quark state.  Coupling to the continuum of meson-baryon final states induces multiple observed resonances from the same bare state.  In EBAC's analysis, all PDG-identified resonances were found to consist of a core state plus meson-baryon components.  (Adapted from Ref.\,\protect\cite{Suzuki:2009nj}.)}}}

\vspace*{8ex}

\end{figure}

Following this line of reasoning, a striking result is agreement between the DSE-computed baryon masses \cite{Roberts:2011cf} and the bare masses employed in modern coupled-channels models of pion-nucleon reactions \cite{Suzuki:2009nj,Gasparyan:2003fp}, see Fig.\,\ref{Fig2} and also Ref.\,\cite{CDRobertsII}.  The Roper resonance is very interesting.  The DSE study \cite{Roberts:2011cf} produces a radial excitation of the nucleon at $1.82\pm0.07\,$GeV.  This state is predominantly a radial excitation of the quark-diquark system, with both the scalar- and axial-vector diquark correlations in their ground state.  Its predicted mass lies precisely at the value determined in the analysis of Ref.\,\cite{Suzuki:2009nj}.  This is significant because for almost 50 years the ``Roper resonance'' has defied understanding.  Discovered in 1963, it appears to be an exact copy of the proton except that its mass is 50\% greater.  The mass was the problem: hitherto it could not be explained by any symmetry-preserving QCD-based tool.  That has now changed.  Combined, see Fig.\,\ref{ebac}, Refs.\,\cite{Roberts:2011cf,Suzuki:2009nj} demonstrate that the Roper resonance is indeed the proton's first radial excitation, and that its mass is far lighter than normal for such an excitation because the Roper obscures its dressed-quark-core with a dense cloud of pions and other mesons.  Such feedback between QCD-based theory and reaction models is critical now and for the foreseeable future, especially since analyses of CLAS data on nucleon-resonance electrocouplings suggest strongly that this structure is typical; i.e., most low-lying $N^\ast$-states can best be understood as an internal quark-core dressed additionally by a meson cloud \cite{MokeevI}.

Additional analysis \cite{CDRobertsII} suggests a fascinating new feature of the Roper.  The nucleon ground state is dominated by the scalar diquark, with a significantly smaller but nevertheless important axial-vector diquark component.  This feature persists in solutions obtained with more sophisticated Faddeev equation kernels (see, e.g., Table~2 in Ref.\,\cite{Cloet:2008re}).  From the perspective of the nucleon's parity partner and its radial excitation, the scalar diquark component of the ground-state nucleon actually appears to be unnaturally large.
One can nevertheless understand the structure of the nucleon.  As with so much else, the composition of the nucleon is intimately connected with DCSB.  In a two-color version of QCD, the scalar diquark is a Goldstone mode, just like the pion.  (This is a long-known result of Pauli-G\"ursey symmetry.)  A ``memory'' of this persists in the three-color theory and is evident in many ways.  Amongst them, through a large value of the canonically normalized Bethe-Salpeter amplitude and hence a strong quark$+$quark$-$diquark coupling within the nucleon.  (A qualitatively identical effect explains the large value of the $\pi N$ coupling constant.) There is no such enhancement mechanism associated with the axial-vector diquark.  Therefore the scalar diquark dominates the nucleon.
The effect on the Roper is striking, with orthogonality of the ground- and excited-states forcing the Roper to be constituted almost entirely from the axial-vector diquark correlation.  One may reasonably expect this to have a material impact on the momentum-dependence of the nucleon-to-Roper transition form factor.

\section{Epilogue}
Dynamical chiral symmetry breaking (DCSB) is a fact in QCD.  It is manifest in dressed-propagators and vertices, and, amongst other things, it is responsible for:
the transformation of the light current-quarks in QCD's Lagrangian into heavy constituent-like quarks, in terms of which order was first brought to the hadron spectrum;
the unnaturally small values of the masses of light-quark pseudoscalar mesons and the $\eta$-$\eta^\prime$ splitting \cite{Bhagwat:2007ha};
the unnaturally strong coupling of pseudoscalar mesons to light-quarks -- $g_{\pi \bar q q} \approx 4.3$;
and the unnaturally strong coupling of pseudoscalar mesons to the lightest baryons -- $g_{\pi \bar N N} \approx 12.8 \approx 3 g_{\pi \bar q q}$.

Herein we have illustrated the dramatic impact that DCSB has upon observables in hadron physics.  A ``smoking gun'' for DCSB is the behaviour of the dressed-quark mass function.  The momentum dependence manifest in Fig.\,\ref{fig:Mp} is an essentially quantum field theoretical effect.  Exposing and elucidating its consequences therefore requires a nonperturbative and symmetry-preserving approach, where the latter means preserving Poincar\'e covariance, chiral and electromagnetic current-conservation, etc.  The Dyson-Schwinger equations (DSEs) provide such a framework.  Experimental and theoretical studies are underway that will identify observable signals of $M(p^2)$ and thereby confirm and explain the mechanism responsible for the vast bulk of visible mass in the Universe.

There are many reasons why this is an exciting time in hadron physics.  We have focused on one.  Namely, through the DSEs, we are positioned to unify phenomena as apparently diverse as: the hadron spectrum; hadron elastic and transition form factors, from small- to large-$Q^2$; and parton distribution functions \cite{Nguyen:2011jy}.  The key is an understanding of both the fundamental origin of nuclear mass and the far-reaching consequences of the mechanism responsible; namely, DCSB.  These things might lead us to an explanation of confinement, the phenomenon that makes nonperturbative QCD the most interesting piece of the Standard Model.

\begin{acknowledgements}
I am grateful for the hospitality and efficiency of the staff at SMU involved in support of \emph{Applications of light-cone coordinates to highly relativistic systems -- Light Cone 2011}; and to the organisers for their kindness and the financial support that enabled my participation.
This contribution was drawn from collaborations and discussions with A.~Bashir, S.\,J.~Brodsky, L.~Chang, C.~Chen, H.~Chen, I.\,C.~Clo\"et, B.~El-Bennich, X.~Guti\'{e}rrez-Guerrero, R.\,J.~Holt, M.\,A.~Ivanov, Y.-x.~Liu, T.~Nguyen, S.-x.~Qin, H.\,L.\,L.~Roberts, R.~Shrock, P.\,C.~Tandy and D.\,J.~Wilson;
and the work was supported by
U.\,S.\ Department of Energy, Office of Nuclear Physics, contract no.~DE-AC02-06CH11357.
\end{acknowledgements}

%%---------
% BibTeX users please use
%\bibliographystyle{spbasic}

%\input BibBaryon.tex

\end{document}